\documentclass[12pt]{amsart}

\newcommand{\R}{\mathbb R}

\newcommand{\x}{\mathbf x}
\newcommand{\y}{\mathbf y}

\newcommand{\eqdef}{\stackrel{\text{def}}{=}}

\newcommand{\Symb}{Symb}
\newcommand{\Diff}{Diff}

\begin{document}

\title[The Poisson algebra of classical Hamiltonians]
{The Poisson algebra of classical Hamiltonians in field theory and the problem of its quantization}
\author{A. Stoyanovsky}
\email{alexander.stoyanovsky@gmail.com}
\address{Russian State University of Humanities}

\begin{abstract}
We construct the commutative Poisson algebra of classical Hamiltonians in field theory.
We pose the problem of quantization of this Poisson algebra.
We also make some interesting computations in the known quadratic part of the quantum algebra.
\end{abstract}

\maketitle

\section{Introduction}
\subsection{QFT Hamiltonians} A typical example of quantum field theory Hamiltonian is
\begin{equation}
\begin{aligned}{}
\hat H(t)&=\int\frac12\left(\hat\pi(\x)^2+\sum_{j=1}^n\left(\frac{\partial\hat\varphi}{\partial x_j}\right)^2(\x)
+m^2\hat\varphi(\x)^2\right)d\x\\
&+\int\left(\frac1{k!}g(t,\x)\hat\varphi(\x)^k+j(t,\x)\hat\varphi(\x)\right)d\x,\\
\end{aligned}
\end{equation}
where $\x=(x_1,\ldots, x_n)$; $\hat\varphi(\x)=\varphi(\x)$
and $\hat\pi(\x)=-ih\frac{\delta}{\delta\varphi(\x)}$ are the operators satisfying the canonical commutation relations
\begin{equation}
[\hat\varphi(\x),\hat\pi(\x')]=ih\delta(\x-\x'),\ \ [\hat\varphi(\x),\hat\varphi(\x')]=[\hat\pi(\x),\hat\pi(\x')]=0;
\end{equation}
$k$ is a natural number; $m$ is a real number called mass;
and $g(t,\x)$, $j(t,\x)$ are scalar functions called respectively the interaction cutoff
function and the source. Usually operators like (1) are considered purely formally.
In this paper we describe the calculus of the corresponding classical Hamiltonians, and
discuss the problem of developing the calculus of quantum operators like (1)
and giving mathematical sense to them.
In the Introduction we collect informal motivations which led to the statement of this problem.

\subsection{First attempts}
The first difficulty
arises when one puts $g\equiv j\equiv0$, $m=1$, omits the term
$\sum\limits_{j=1}^n\left(\frac{\partial\hat\varphi}{\partial x_j}\right)^2(\x)$ in (1) and
considers the remaining expression
\begin{equation}
\int\frac12\left(\hat\pi(\x)^2+\hat\varphi(\x)^2\right)d\x.
\end{equation}
When trying to take the square of this expression and trying to transfer the operators $\hat\pi(\x)$ to the right
using the relations (2),
one gets the divergent expression $\int\int\delta(\x-\x')^2d\x d\x'$. If one tries to avoid this, then one should
assume that the algebra in question should contain the operators like
\begin{equation}
\int\frac12\left(\hat\varphi(\x)\hat\pi(\x)+\hat\pi(\x)\hat\varphi(\x)\right)d\x.
\end{equation}
The first assumption made by the author (see, for example, [2]) was that the required algebra coincides with the
infinite dimensional analog of the Weyl--Moyal algebra which contains expressions like (4). However, it is not difficult
to see that the square of expression (3) does not belong to the Weyl--Moyal algebra as well, so it also does not
suit for our purposes.

\subsection{The finite dimensional analogy}
The next important observation occurs if one considers the finite dimensional situation. Consider the Schrodinger
equation
\begin{equation}
ih\frac{\partial\psi}{\partial t}=\hat H\psi,
\end{equation}
where $\psi=\psi(t,q)$, $q=(q_1,\ldots,q_N)$, $H=H(t,p_i,q_i)$ is the classical Hamiltonian, and
$\hat H=H(t,-ih\partial/\partial q_i,q_i)$ is the quantum Hamiltonian. Here one meets the problem of the choice of
the ordering of the operators $q_i$ and $-ih\partial/\partial q_j$ in the quantum Hamiltonian; assume for a moment that
all $-ih\partial/\partial q_j$ are put to the right of $q_i$. Let us substitute into equation (5) a quasiclassical
solution of the form
\begin{equation}
\psi(t,q)=a(t,q)e^{iS(t,q)/h},
\end{equation}
and consider equation (5) up to $o(h)$. Equating the principal terms, one obtains the Ha\-mil\-ton--Jac\-obi equation
for $S(t,q)$:
\begin{equation}
\frac{\partial S}{\partial t}+H(t,\frac{\partial S}{\partial q_i},q_i)=0.
\end{equation}
Further, equating the terms before $h$, one gets the transport equation for the amplitude $a(t,q)$:
\begin{equation}
\begin{aligned}{}
\frac{\partial a}{\partial t}&+\sum_i
\frac{\partial a}{\partial q_i}H_{p_i}(t,
\frac{\partial S}{\partial q_1},\ldots,\frac{\partial S}{\partial q_N},q_1,\ldots,q_N)\\
&+\frac a2\sum_{i,j}H_{p_ip_j}(t,
\frac{\partial S}{\partial q_1},\ldots,\frac{\partial S}{\partial q_N},q_1,\ldots,q_N)
\frac{\partial^2S}{\partial q_i\partial q_j}=0.
\end{aligned}
\end{equation}
It is well known that a solution to this equation in the particular case $\frac12\sum_iH_{p_iq_i}\equiv0$ (this condition
is related to ``non-canonicity'' of correspondence between classical and quantum Hamiltonians)
is given by the formula
\begin{equation}
a(t,q(t))=a_0(q(0))\frac1{\sqrt{\det\left(\frac{\partial q_i(t)}{\partial q_j(0)}\right)}},
\end{equation}
where $(p_i(t),q_i(t))$ is a characteristic, i.~e., a solution of the classical Hamilton equations;
$p_i=\partial S/\partial q_i$. In the infinite dimensional case the Hamilton--Jacobi equation and its solution
have sense (see [3] and references therein), but formula (9) does not make sense.
This leads to an assumption that the wave function $\psi(q)$ (and its infinite dimensional analog) changes under
the diffeomorphisms of the $q$-space not like a function but like a {\it half-form}, i.~e. it is actually an
expression of the form $\psi(q)(dq_1\ldots dq_N)^{1/2}$.

\subsection{Half-forms on an infinite dimensional space} Returning to the infinite dimensional case, we arrive at the
following problem [1]: define a space of (distribution) half-forms on infinite dimensional space of functions
$\varphi(\x)$ with the following properties:

1) it is acted on by a group of diffeomorphisms of the space of functions;

2) it is acted on by a ring of differential operators like (1);

3) for two half-forms $\Psi_1, \Psi_2$ the number $\langle\Psi_1|\Psi_2\rangle$ is ``usually'' defined which can be
finite or infinite; it is linear with respect to $\Psi_1$, anti-linear with respect to $\Psi_2$, positive definite, and
diffeomorphism invariant.

Let us comment on requirement (1), in addition to the comments in 1.3.
The configuration spaces met in quantum field theory can be more complicated than
just the space of functions; they can be infinite dimensional manifolds (for example, in non-spinor
electrodynamics or in pure
Yang--Mills theory it is the quotient space of the space of connections by the gauge group). Hence one should be
able to define half-forms also for open subsets of a vector space and for manifolds, and one should have action of
diffeomorphisms on them. Here one can also understand why the Weyl--Moyal calculus does not suit us:
it is not invariant under diffeomorphisms of the space of functions.

For example, it is clear that expression (4) makes sense if we consider it as
a Lie derivative ``acting on half-forms'' instead of usual
functionals.

\subsection{The free Klein--Gordon field} The final remark that we want to make in this Introduction is concerned
with the particular case $g\equiv j\equiv0$ of operator (1). The corresponding field theory with the quadratic
Hamiltonian (1) is called the {\it free
Klein--Gordon field}, since in this case the classical field equation coincides with the Klein--Gordon equation
\begin{equation}
\Box\varphi-m^2\varphi=-\frac{\partial^2\varphi}{\partial t^2}+\sum_{j=1}^n\frac{\partial^2\varphi}{\partial x_j^2}
-m^2\varphi=0.
\end{equation}
One expects that the Schrodinger equation with the quadratic Hamiltonian must be solved exactly, as in the finite
dimensional case. This means that the Lie algebra of classical quadratic Hamiltonians with respect to the Poisson
bracket should act on the imagined space of half-forms, and hence it should belong to the ring of
differential operators. This Lie algebra is an infinite dimensional symplectic Lie algebra. The evolution operators
of the Klein--Gordon equation from one (possibly curved) Cauchy space-like surface to another should belong to
the corresponding symplectic group, cf. [3]. These operators are naturally continuous linear symplectic operators on the
symplectic space $S\oplus S$, where $S$ is the Schwartz space (not on $S\oplus S'$!).
Hence we come to a conclusion that the quadratic
part of our algebra of differential operators should coincide with the infinite dimensional symplectic Lie algebra of this
topological vector space.

\subsection{Plan of the paper} \S2 contains the construction of the Poisson algebra of classical
Hamiltonians. In \S3 we state the problem of its quantum deformation.
\S4 is devoted to a curious computation in the quantum algebra.

The author is grateful to M. Finkelberg for stimulating discussions.

\section{Construction of the Poisson algebra of classical Hamiltonians}

Let $S$ be a real nuclear separable topological vector space [4]. For concreteness,
we will mostly consider the case when $S$ is the Schwartz space of functions $\varphi(\x)$ on $\R^n$,
but all the results hold in the general case.

We want to construct the algebra of ``differential operators acting
on half-forms on $S$''. Let us first construct the corresponding commutative Poisson
``algebra of symbols''.

Recall that a functional $\Phi(\varphi(\cdot))$ on $S$ is called {\it continuously
differentiable} if its weak differential or {\it variation}
\begin{equation}
\delta\Phi(\varphi,\delta\varphi)\eqdef
\lim_{\varepsilon\to0}\frac{\Phi(\varphi+\varepsilon\cdot\delta\varphi)-\Phi(\varphi)}
{\varepsilon}
\end{equation}
exists and is a continuous functional of $(\varphi,\delta\varphi)\in S\times S$. Such $\delta\Phi$ is automatically
linear in $\delta\varphi$, so we can consider it as a distribution in $\x$ denoted $\frac{\delta\Phi}{\delta\varphi(\x)}$:
\begin{equation}
\delta\Phi(\varphi,\delta\varphi)\eqdef\int\frac{\delta\Phi}{\delta\varphi(\x)}\delta\varphi(\x)d\x.
\end{equation}
This distribution $\frac{\delta\Phi}{\delta\varphi(\x)}(\varphi)$ is called the {\it variational derivative} of $\Phi$
at $\varphi\in S$ at the point $\x$. It is the analog of partial derivatives.

Repeatedly differentiating $\delta\Phi$ with respect to $\varphi$, we obtain the definitions of second variation
$\delta^2\Phi(\varphi,\delta_1\varphi,\delta_2\varphi)$,
third variation, etc. We require them to be continuous functionals of
$(\varphi,\delta_1\varphi,\delta_2\varphi)\in S\times S\times S$, etc. The second variation is bilinear and symmetric in
$\delta_1\varphi$ and $\delta_2\varphi$.
The second variational derivative is, by definition, the symmetric distribution in $\x$ and $\y\in\R^n$ defined
(by the Schwartz kernel theorem [4]) by the equality
\begin{equation}
\delta^2\Phi(\varphi,\delta_1\varphi,\delta_2\varphi)=\int\frac{\delta^2\Phi}{\delta\varphi(\x)\delta\varphi(\y)}
\delta_1\varphi(\x)\delta_2\varphi(\y)d\x d\y.
\end{equation}

Let us now describe symbols, i.~e., functions on the cotangent bundle to $S$. To this end, we need an additional
structure on $S$. This structure can be given by a continuous scalar product on $S$ defined up to
multiplication by a positive smooth function of polynomial growth (together with all derivatives) $\rho(\x)$:
\begin{equation}
(\varphi_1,\varphi_2)=\int\varphi_1(\x)\varphi_2(\x)\rho(\x)d\x.
\end{equation}
The function $\rho(\x)$ can be arbitrary; what follows does not depend on it.

This scalar product enables us to identify $S$ with its image in the dual space $S'$ of tempered distributions.
In what follows we will need only the existence of this subspace in $S'$.

Let us first describe the cotangent bundle to $S$ that we shall use. It is just the product $S\times S$ considered
as a subset of $S\times S'$.

{\bf Definition.}
A functional $\Phi(\varphi,\pi)$ on the cotangent bundle $S\times S$ is called a {\it classical Hamiltonian}
or a {\it symbol} if it is infinitely differentiable, and its first variational derivatives
$\left(\frac{\delta\Phi}{\delta\varphi(\x)},\frac{\delta\Phi}{\delta\pi(\y)}\right)$
are in fact smooth rapidly decreasing functions in $\x$ and in $\y$, i.~e. belong to
$S\oplus S\subset S'\oplus S'$ for any $(\varphi,\pi)\in S\times S$; moreover, $\delta\Phi$ is infinitely differentiable
as an $S\oplus S$-valued functional of $(\varphi,\pi)\in S\times S$.

Clearly, the symbols form a commutative algebra. The Poisson bracket is given by the usual formula
\begin{equation}
\{\Phi_1,\Phi_2\}=\int\left(\frac{\delta \Phi_1}{\delta\pi(\x)}
\frac{\delta \Phi_2}{\delta\varphi(\x)}-\frac{\delta \Phi_1}{\delta\varphi(\x)}
\frac{\delta \Phi_2}{\delta\pi(\x)}\right)d\x.
\end{equation}
It is interesting to note that the Poisson bracket of two symbols is again a symbol (it is a direct check).
This Poisson bracket satisfies the Leibnitz rule with respect to multiplication.
It also satisfies the Jacobi identity (it is also a direct check).

A {\it polynomial symbol} is a symbol which is a polynomial functional of $(\varphi,\pi)\in S\times S$.
Denote the Poisson algebra of polynomial symbols by $\Symb$. Introduce the grading on this algebra by degree
{\it with respect to~$\pi$}:
\begin{equation}
\Symb=\bigoplus_{k\ge0}\Symb_k.
\end{equation}

\section{The problem of quantization of the constructed Poisson algebra}

{\bf Problem.} Construct a family of filtered vector spaces
\begin{equation}
\Diff_h=\bigcup_{k\ge 0}F_k\Diff_h
\end{equation}
depending on the parameter $h$, with associative multiplication
\begin{equation}
\cdot:\Diff_h\times\Diff_h\to\Diff_h,\ \ F_k\Diff_h\cdot F_l\Diff_h\subset F_{k+l}\Diff_h,
\end{equation}
so that for $h=0$ the algebra $\Diff_0$ coincides with $\Symb$;
moreover, for any $h$ there are fixed isomorphisms
\begin{equation}
Gr_k\Diff_h\eqdef F_k\Diff_h/F_{k-1}\Diff_h\simeq\Symb_k,\ \ k\ge 0,
\end{equation}
and the multiplication in the associated graded algebra
\begin{equation}
Gr\Diff_h=\bigoplus_{k\ge0}Gr_k\Diff_h
\end{equation}
coincides with multiplication in $\Symb$. The commutator in $\Diff_h$ up to $o(h)$ should
coincide with the Poisson bracket in $\Symb$ multiplied by $-ih$.

\section{A computation in the quantum algebra}

Assume that $n=1$; let us write $x$ instead of $\x$. Let us compute the commutator
\begin{equation}
\left[\int f(x)\hat\varphi(x)\hat\varphi'(x)dx,\int g(y)\hat\pi(y)^2dy\right],
\end{equation}
where $f,g\in S$. Of course, due to what was said in the Introduction, this is equivalent to computation of the
Poisson bracket of two quadratic symbols (or a commutator in the symplectic Lie algebra).
Below we shall omit $\int f(x)g(y)dxdy$ in our computations.
We have, under the integral, by computing the commutator in two ways (the first way by applying the Leibnitz rule
to the left term, and the second way by applying this rule to the right term):
\begin{equation}
\begin{aligned}{}
&\frac1{ih}[\hat\varphi(x)\hat\varphi'(x),\hat\pi(y)^2]\\
&=2\hat\pi(x)\hat\varphi'(x)\delta(x-y)+2\hat\varphi(x)\hat\pi'(x)\delta(x-y)\\
&+2\hat\varphi(x)\hat\pi(x)\delta'(x-y)\\
&=(\hat\varphi'(x)\hat\pi(x)+\hat\pi(x)\hat\varphi'(x)+\hat\varphi(x)\hat\pi'(x)+\hat\pi'(x)\hat\varphi(x))\delta(x-y)\\
&+(\hat\varphi(x)\hat\pi(x)+\hat\pi(x)\hat\varphi(x))\delta'(x-y).
\end{aligned}
\end{equation}
Subtracting the right hand side of the latter equality from the left hand side, we obtain
\begin{equation}
([\hat\pi(x),\hat\varphi'(x)]+[\hat\varphi(x),\hat\pi'(x)])\delta(x-y)
+[\hat\varphi(x),\hat\pi(x)]\delta'(x-y)=0,
\end{equation}
whence, replacing
\begin{equation}
\begin{aligned}{}
[\hat\varphi(x),\hat\pi(x)]&=ih\delta(x-y)|_{y=x}=ih\delta(0),\\
[\hat\pi(x),\hat\varphi'(x)]&=-ih\delta'(x-y)|_{y=x}=-ih\delta'(0),\\
[\hat\varphi(x),\hat\pi'(x)]&=-ih\delta'(x-y)|_{y=x}=-ih\delta'(0)
\end{aligned}
\end{equation}
and putting $y=0$, we obtain a curious identity
\begin{equation}
\delta(0)\delta'(x)=2\delta'(0)\delta(x).
\end{equation}
This identity can be also obtained by formally differentiating the equality
\begin{equation}
\delta(x)^2=\delta(0)\delta(x).
\end{equation}

\end{document}